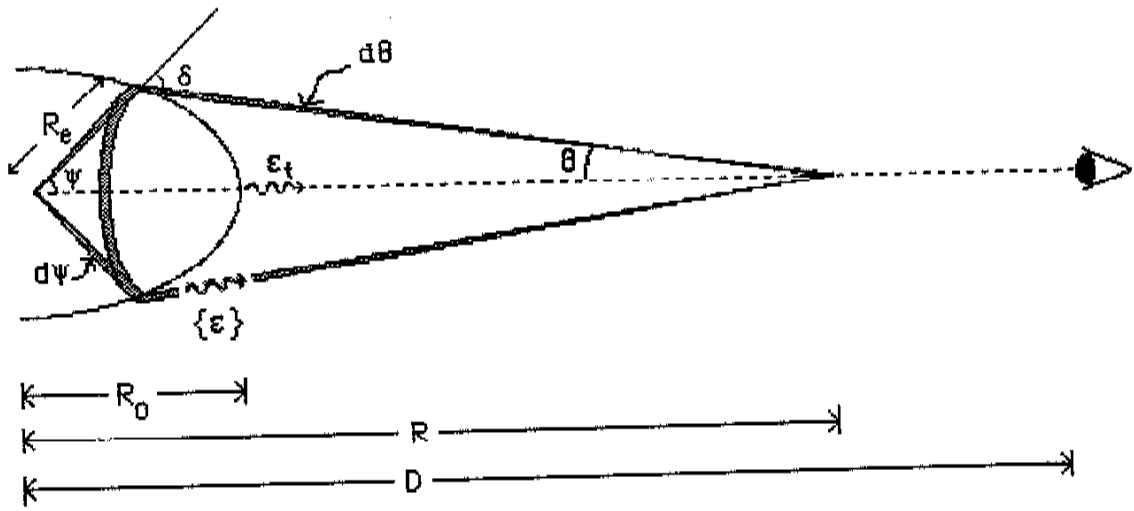

**Fig. 1**

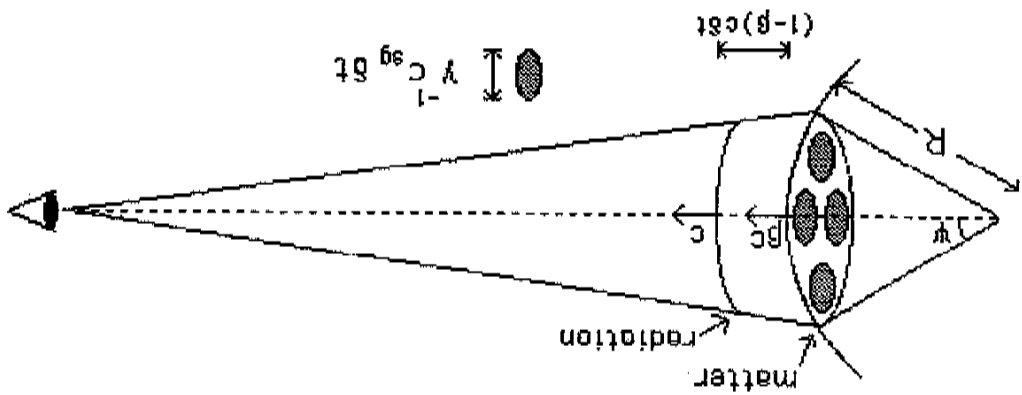

Fig. 3

# EMPIRICAL CONSTRAINTS ON SOURCE PROPERTIES AND HOST GALAXIES OF COSMOLOGICAL GAMMA-RAY BURSTS

Eric Woods and Abraham Loeb

Astronomy Department, Harvard University, 60 Garden St., Cambridge MA 02138

## ABSTRACT

We discuss several constraints on the properties of $\gamma$-ray bursts (GRB) at cosmological distances. First we use the requirement that burst sources must be optically thin to a test photon for the process $\gamma + \gamma \to e^+ + e^-$ in order to produce the observed nonthermal spectra. In particular, we derive probability distributions for the minimum Lorentz expansion factor $\gamma_{\min}$, the radiation energy $E_\gamma$, the maximum baryonic mass $M_{\max}$, and the maximum surrounding gas density $n_{\max}$ in the events, based on 254 events from the second BATSE catalog. In the case where the GRB spectrum cuts off at the highest observed energies ($\sim 100$ MeV), we obtain the mean values $\langle \gamma_{\min} \rangle = 90$, $\langle E_\gamma \rangle = 4 \times 10^{51} h^{-2}$ ergs, and $\langle M_{\max} \rangle = 3 \times 10^{-5} \xi^{-1} M_\odot$, where $\xi$ is the fraction of the total energy which is converted to $\gamma$-rays. The distribution of burst energies ends at about $10^{53}$ ergs, close to the binding energy of a neutron star.

Secondly, the time variabilities of the bursts in the BATSE catalog are used to place an upper bound $R_{\max}$ on the curvature radius of the emitting surfaces in the events. This is based on the requirement that the emitting region seen by the observer must be sufficiently small to produce the observed variability without violating causality. Using the 64-msec resolution of BATSE, we find that a significant number of bursts have $R_{\max} \approx 10^{13} (\gamma/10^2)^2$ cm, where $\gamma$ is the Lorentz factor of the expansion. This limit should become stricter with finer time resolution.

Finally, we discuss the association of cosmological GRB with galaxies. We consider eight bright and well-localized bursts detected by the Pioneer Venus Orbiter, whose positional error boxes contain no bright galaxies. Assuming that burst events occur in galaxies, we place upper limits on the luminosities of the host galaxies. Using the local luminosity function of galaxies, we calculate the probability for not seeing the GRB host galaxy. This probability tends to increase as the width of the GRB luminosity function increases. However, the allowed width of the GRB luminosity function is restricted by the burst peak flux distribution.

*Subject headings:* cosmology: observations–gamma rays: bursts

# 1. INTRODUCTION

Gamma-ray bursts (GRB) have eluded a definitive explanation for more than twenty years. Hundreds of models have been proposed in the past, most of these involving Galactic sources (cf. Nemiroff 1994). More recently, however, the Burst And Transient Source Experiment (BATSE) on board the Compton Gamma-Ray Observatory (GRO) has shown that the burst population is highly isotropic (Meegan et al. 1993; Briggs et al. 1993). This result is inconsistent with a Galactic disk population of sources, but suggests that the bursts occur either in an extended Galactic halo or at cosmological distances (Mao & Paczyński 1992; Hakkila et al. 1994). There is also a deficiency of faint bursts relative to a uniform distribution in Euclidean space, which can be naturally explained if the faint bursts are at high redshifts (Mao & Paczyński 1992; Piran 1992; Dermer 1992). A cosmological redshift is also implied by the suggestion that dim bursts are on average longer than bright bursts (Norris et al. 1993). Since little is known about the origin of cosmological GRB at this stage, it is important to extract the maximum amount of information on the burst sources from the existing data. In this paper we discuss general empirical constraints on cosmological $\gamma$-ray burst models, and use the second BATSE catalog to quantify these constraints for the entire burst population. Some of our constraints were briefly discussed in a previous paper on the subject (Woods & Loeb 1994).

First, let us consider qualitatively the physics behind the various constraints derived in this work. A typical burst located at a distance of order the Hubble radius ($c/H_0 \sim 3h^{-1}$ Gpc) must release a large amount of energy ($\sim 10^{51}$ ergs) to produce the observed fluence. Combining this with the small source size ($\lesssim 3 \times 10^{10}$ cm) associated with the typical variability timescale of $\sim 1$ second, one obtains an optical depth to pair production ($\gamma + \gamma \to e^+ + e^-$) of order $\sim 10^{11}$. Under these circumstances, the photon distribution should be thermalized, in conflict with the observed nonthermal spectra of GRB ($\nu I_\nu \approx$ const between 0.1 and 100 MeV). The thermalization problem is avoided if the the emitting photosphere of the burst source expands relativistically with a Lorentz factor $\gamma \gg 1$ (Goodman 1986; Paczyński 1986). The emission is then beamed into cones of half angle $\sim \gamma^{-1}$, so that the center-of-mass momentum in photon-photon collisions is lowered below the pair-production threshold, and the characteristic photon paths cross only at large distances where the scattering probability is low. For typical burst parameters, the requirement for an optically thin source translates to the bound $\gamma \gtrsim 10^2$ for emission from a fixed radius (Fenimore, Epstein, & Ho 1993). For emission from an expanding shell, this constraint is weakened considerably, so that $\gamma \lesssim 10^2$ is allowed.

This limit on the Lorentz factor may be used to place constraints on other burst properties. The minimum $\gamma$ can be used to set an upper bound on the baryonic mass $M$ participating in the GRB event (Shemi & Piran 1990). This constraint is obtained through the relation $M = E_\gamma/\xi\gamma c^2$, where $E_\gamma$ is the total $\gamma$-ray energy released, $c$ is the speed of light, and $\xi$ is the fraction of the total energy released in $\gamma$-rays. Based on this upper bound on the number of baryons in the burst, it is possible to limit the ambient gas density which is swept by the expanding photosphere of the source. We define this bound as $n_{\max}$.

Until recently all of the above limits were evaluated only roughly for "average" values of the burst parameters. In an earlier publication, the first BATSE catalog allowed us to get a first statistical evaluation of these constraints for the entire burst population, assuming emission from a fixed radius (Woods & Loeb 1994). In §2.1 and §3.1 we assume the more physical case of emission from an expanding shell, and derive these constraints for an even larger sample of bursts, using the second BATSE catalog and its associated data set of GRB light curves. Moreover, we include two additional constraints as described below.

Some models require that the main part of the emission take place at a radius larger than some minimum radius. In the fireball model of Mészáros, Laguna, & Rees (1993), a burst is produced when the expanding fireball decelerates due to the surrounding medium. For typical burst energies ($E \sim 10^{51}$ erg) and typical interstellar densities ($\sim 1$ cm$^{-3}$), the fireball is not decelerated appreciably until it has reached a radius $R \sim 10^{16} (\gamma/10^3)^{-2/3}$ cm. If an empirical *upper* bound can be placed on the radius at which the burst is produced, then such models can be constrained.

An upper bound of this type can be obtained from the variability timescales of the bursts in the second BATSE catalog. Let us assume that the emission takes place at a radius $R$ while the emitting surface expands with speed $\beta c$ and corresponding Lorentz factor $\gamma = (1 - \beta^2)^{-1/2}$, and that signals propagate in the rest frame of the expanding material at speeds $\sim c_{\rm sg}$. A burst which exhibits variability $\delta C/C$ in its count rate $C$ on a comoving timescale $\delta t'$ must have originated from a region smaller than the signal-travel time $\ell \sim c_{\rm sg} \delta t'$ due to causality. The observed timescale corresponding to $\delta t'$ is $\delta t_o = \gamma(1-\beta)\delta t' \approx (2\gamma)^{-1}\delta t'$, where the first factor $\gamma$ is due to the Lorentz transformation from the comoving frame to the lab frame, and $(1-\beta)$ accounts for the velocity difference between the expanding material and the photons. Hence, $\ell \sim 2\gamma c_{\rm sg} \delta t_o$. An area larger than $\sim \ell^2$ will in general contain many regions which are causally disconnected from one another on the observed timescale $\delta t_o$. These regions act as independent emission sites, and by Poisson statistics produce variations $\delta C/C \sim N^{-1/2}$, where $N$ is the number of emitting regions of size $\lesssim \ell$ that are effectively contributing to the observed signal. Since relativistic beaming limits the emission to an angle $\theta \sim 1/\gamma$, the effective emitting area of the source is $\sim \theta^2 R^2$, where $R$ is the curvature radius of the (not necessarily spherical) emitting surface. Therefore, the minimum number of causally disconnected regions at the emission photosphere is given by $N \gtrsim \theta^2 R^2/\ell^2 \sim (R/2\gamma^2 c_{\rm sg} \delta t_o)^2$, and the resulting fluctuation amplitude is $\delta C/C \lesssim 2\gamma^2 c_{\rm sg} \delta t_o/R$. This condition can in turn be used to find the maximum curvature radius $R_{\rm max} = 2\gamma^2 c_{\rm sg} \delta t_o/(\delta C/C)$ which could produce the observed variabilities $\delta C/C$ in each individual burst. We discuss this constraint in detail in §2.2 and §3.2.

As the last empirical constraint, we consider the potential association of GRB with galaxies. Searches for GRB counterparts at other wavelengths have all turned up empty-handed (Schaefer 1993), and there is doubt as to whether extragalactic bursts could even be associated with galaxies (Schaefer 1992; Fenimore et al. 1993). Any model which places burst sources inside of galaxies must be consistent with observational constraints on the galaxies near GRB positions on the sky. If bursts occur outside of galaxies, one needs to turn to exotic emission mechanisms (e.g. cosmic strings; Paczyński 1988; Babul, Paczyński, & Spergel 1987) or postulate the existence of an as yet unobserved population of neutron stars in the intergalactic space. In the latter case, the neutron stars could have been ejected from galaxies with large kick velocities (Narayan et al. 1992).

The positional error boxes associated with eight well-localized bursts were searched for galaxies, and none were found down to very faint magnitudes $B \lesssim 24$. This null result was used by Schaefer (1992) to derive a lower bound on the distance to a burst, assuming the burst is associated with a galaxy whose luminosity is comparable to that of M31. Fenimore et al. (1993) used Schaefer's results, together with the combined BATSE-PVO peak flux distribution, to obtain limits on the absolute magnitudes of the GRB host galaxies. In §4, we derive more restrictive constraints from the same data. In particular, the galaxy luminosity function is used to set constraints on the width of the GRB luminosity function. Obviously, if the luminosity function of GRB is sufficiently broad, it should allow for some bright bursts to be at large distances, where their host galaxies are too faint to be seen. Thus, the assumption that most GRB occur inside galaxies can be used to set a lower limit on the width of the GRB luminosity function.

A summary and discussion of all of the above results is given in §5. Throughout the paper we assume a Hubble constant $H_0 = 75$ km sec$^{-1}$ Mpc$^{-1}$.

## 2. CONSTRAINTS ON SOURCE PROPERTIES: THEORY

In this section we derive constraints on properties of the burst sources in terms of quantities measured by the BATSE experiment. The distances to the individual bursts are calibrated according to their peak flux. This distance estimator provides an excellent fit to the number count statistics in standard cosmologies (cf. Fig. 4a, later).

### 2.1 Pair Production Optical Depth of Cosmological Bursts

We first calculate the optical depth for pair production by photon-photon collisions near a cosmological source. Because of the conservation of relativistic energy-momentum, a test photon of energy $\epsilon_t$ can only produce an $e^+e^-$ pair in a collision with a photon whose energy $\epsilon$ is greater than $\epsilon_{\rm th} \equiv 2m_e^2 c^4/\epsilon_t(1-\cos\theta)$, where $m_e$ is the electron rest mass and $\theta$ is the angle between the photon trajectories. The cross section for collisions above this threshold is

$$\sigma(\epsilon_t, \epsilon, \theta) = \frac{3}{16}\sigma_{\rm T}(1-v^2)\left[(3-v^4)\ln\frac{(1+v)}{(1-v)} - 2v(2-v^2)\right], \qquad (1)$$

where $\sigma_{\rm T}$ is the Thomson cross section and $v = [1-(\epsilon_{\rm th}/\epsilon)]^{1/2}$ is the center-of-mass speed of the outgoing pair particles in units of the speed of light (e.g. Berestetskii, Lifshitz, & Pitaevskii 1982).

Suppose now that a large amount of energy is released at a point in space, in the form of relativistically expanding matter (baryons, $e^+e^-$ pairs, or even exotic particles). At some distance from the source, the matter begins to produce $\gamma$- rays isotropically in its rest frame; the emission continues until the expanding shell has reached a radius $R_0$ (see Fig. 1). The mechanism of $\gamma$-ray production is not important; it could be the interaction of a baryon-pair fireball with the surrounding medium (Mészáros et al. 1993 and references therein), or collisions between shells (Rees & Mészáros 1994). If the matter expands with a Lorentz factor $\gamma \gg 1$, the emission is strongly beamed, and the observer will see photons only from a small patch of the emitting surface. Thus, for our purposes the emitting surface need not be spherical; it could be a collimated jet with an opening angle $> \gamma^{-1}$ and a curvature radius $R_0$.

**Fig. 1:** Geometry for the calculation of the pair-production optical depth $\tau$ for an expanding source. The shaded area represents the contribution to $\tau$ at a distance $R$ from the source from a ring of angle $\psi$ and angular width $d\psi$. The surface shown is the locus of points from which photons originate whose trajectories intersect that of the test photon at the instant it reaches a distance $R$ (Fenimore et al. 1993).

Consider a test photon of energy $\epsilon_t$ which is emitted at the end of the emission, at a distance $R_0$ from the center of the explosion, and which then propagates precisely along the line of sight. When the test photon has reached a distance $R$ from the center, only photons which arrive at this point at precisely the same instant can collide with the test photon to produce pairs. The locus of points from which these photons arrive defines a surface $R_e(\psi)$, where $R_e$ is the emission radius and $\psi$ is the angle the emission radius vector makes with the line of sight at the center of the explosion (see Fig. 1). This surface is given by

$$\frac{(R_e^2 - 2RR_e\cos\psi + R^2)^{1/2}}{c} + \frac{R_e}{\beta c} = \frac{R_0}{\beta c} + \frac{R - R_0}{c}, \tag{2}$$

where $\beta = (1 - \gamma^{-2})^{1/2}$ to be the expansion speed in units of the speed of light. The solution to equation (2) is

$$\begin{aligned}R_e = &\gamma^2[R_0 + \beta(R - R_0) - \beta^2 R\cos\psi] \\ &- \{\gamma^4[R_0 + \beta(R - R_0) - \beta^2 R\cos\psi]^2 - \gamma^2(1-\beta)[(1-\beta)R_0^2 + 2\beta R R_0]\}^{1/2}.\end{aligned} \tag{3}$$

We assume that the emission is steady over a timescale $\sim R_0/c$. The optical depth $\tau$ is obtained by integrating the collision probability from $R_0$ out to infinity. At each point in the test photon's path, one must also integrate over the photon energy distribution, and over the contributions from different collision angles:

$$\tau(\epsilon_t) = \frac{1}{c}\int_{R_0}^{\infty} dR \int_{\epsilon_{th}}^{\infty} d\epsilon \int dA \frac{d\phi(\epsilon, R)}{dA} \sigma(\epsilon_t, \epsilon, \theta), \tag{4}$$

where $d\phi(\epsilon, R)$ is the contribution to the photon flux per unit energy, at a radius $R$, from an area element $dA$ on the emitting surface. Photons emitted from $dA$ collide with the test photon at an angle $\theta$.

Figure 1 shows the contribution at a distance $R$ to the optical depth from photons originating in a ring which subtends angles in the range $(\psi, \psi + d\psi)$ at the center of the explosion. We define $\delta$ to be the angle between the photon trajectory and the radial direction in the observer's frame. Note that the observed energies of photons from this ring are blueshifted from their rest-frame energies by a factor $[\gamma(1 - \beta \cos \delta)]^{-1}$. However, for $\gamma \gg 1$, the surface "seen" by the test photon at any given instant always covers an angle $\sim \gamma^{-1}$, due to the restriction imposed by equation (2), and the effect of relativistic beaming (cf. Fenimore et al. 1993). Thus, the range of blueshifts contributing to the total flux is the same at all distances $R$ from the source, so we may make the approximation that the same spectrum is observed at all radii. A detailed calculation shows that this approximation yields nearly the same results as the case where the variation of the spectrum with radius is taken into account.

The photon flux per unit emitting area at a distance $R$ from the source is given by

$$\frac{d\phi(\epsilon, R)}{dA} = \frac{\phi(\epsilon, R)}{J(R)[\gamma(1 - \beta \cos \delta)]^2}. \tag{5}$$

The two factors of $\gamma(1 - \beta \cos \delta)$ account for the relativistic beaming due to the Lorentz transformation of solid angles (cf. Rybicki & Lightman 1979). We have defined the normalization $J(R)$ such that the integral of $d\phi/dA$ over the entire contributing surface yields the total flux $\phi(\epsilon, R)$,

$$J(R) \equiv \int \frac{dA}{[\gamma(1 - \beta \cos \delta)]^2} = \int_0^{\psi_{\max}} \frac{2\pi R_e^2 [1 + (R_e'/R_e)^2]^{1/2} \sin \psi \, d\psi}{[\gamma(1 - \beta \cos \delta)]^2}, \tag{6}$$

where $R_e$ is given by equation (3), $R_e' \equiv dR_e/d\psi$, and $\psi_{\max}(R)$ is the maximum angle which contributes at a distance $R$, i.e., the angle such that the emitted photon's trajectory is tangent to the surface (see Fig. 1). The flux per unit energy at a distance $R$ is given in terms of the observed flux per unit energy $\phi_o(\epsilon)$ by

$$\phi(\epsilon, R) = \frac{D^2}{R^2} \frac{1}{(1+z)^2} \phi_o\left(\frac{\epsilon}{1+z}\right), \tag{7}$$

for a source at a cosmological redshift $z$, and a corresponding luminosity distance $D(z)$ (e.g. Weinberg 1972). The additional factor of $(1+z)^{-2}$ is due to the fact that we are considering the number flux of photons per unit energy rather than their total energy flux. Equations (1)-(7) then provide the total optical depth of a GRB source, with a bulk expansion Lorentz factor $\gamma$, to a test photon with energy $\epsilon_t$:

$$\tau(\gamma, \epsilon_t) = \frac{D^2}{c(1+z)^2} \int_{R_0}^{\infty} \frac{dR}{R^2 J(R)} \int_{\epsilon_{th}}^{\infty} d\epsilon \, \phi_o\left(\frac{\epsilon}{1+z}\right)$$
$$\times \int_0^{\psi_{\max}} d\psi \, \sin \psi \, \frac{2\pi R_e^2 [1 + (R_e'/R_e)^2]^{1/2}}{[\gamma(1 - \beta \cos \delta)]^2} \, \sigma(\epsilon_t, \epsilon, \theta). \tag{8}$$

Note that the cross-section $\sigma$ depends on the collision angle $\theta$, while the relativistic emission probability depends on the emission angle $\delta$. Inspection of the triangle in figure 1 yields the relations $\delta = \psi + \theta$, $\tan\theta = R_0 \sin\psi/(R - R_0 \cos\psi)$, and $\sin\theta = (R_0/R)\sin\delta$.

The burst spectrum can usually be approximated by a power law,

$$\phi_o(\epsilon) = k\epsilon^{-\alpha}, \qquad \epsilon_1 < \epsilon < \epsilon_2, \qquad (9)$$

Most burst spectra seem to be more or less featureless continua with roughly equal energy flux per logarithmic energy interval, i.e., $\alpha = 2$ in equation (9). The spectra are highly nonthermal, with no sign of a significant decline at any value of $\epsilon_2$, even out as far as $\epsilon_2 = 100$ MeV. For an expanding source with $\alpha = 2$, the optical depth $\tau$ is quite sensitive to the energy cutoff $\epsilon_2$. Fenimore et al. (1993) showed that for large enough values of $\gamma$, all of the photons will collide with the test photon at angles less than the critical angle $\theta_{\rm cr} \approx 2m_e c^2/(\epsilon_{\rm t}\epsilon_2)^{1/2}$, meaning that the center-of-mass energy will always be below the threshold value $\epsilon_{\rm th}$ and no pairs will be produced. Since burst spectra have been observed to extend out to energies of order 100 MeV, we take a test photon energy $\epsilon_{\rm t} = 100(1+z)$ MeV, and will consider two extreme cases: the case where the spectrum extends out to infinite energies ($\epsilon_2 \to \infty$), and the case where the spectrum cuts off at an energy of the order of the test photon energy ($\epsilon_2 = 100$ MeV). Note that since most of the contribution to $\tau$ comes from near the source, we are justified in integrating out to infinite radius.

Implicitly built into this calculation is the assumption that the emission is steady on timescales $\sim R_0/c$. This corresponds to an observed timescale $\delta t_o \sim (1-\beta)R_0/c \approx R_0/(2c\gamma^2)$, where the factor $(1-\beta)$ accounts for the shortening of the duration due to the source's expansion toward the observer. This is of the same order as the geometric time delay between the arrival of photons along the line of sight and photons coming in from the edge of the observed emission region, at angles $\psi \sim \gamma^{-1}$. Since all signals are smeared out by this delay, there is no way to test this assumption empirically. However, the assumption is natural since plausible physical processes that convert the kinetic energy of the fireball to $\gamma$-rays after a time $R_0/c$ (e.g. deceleration by the ambient gas or shell collisions) should typically last a time $\gtrsim R_0/c$.

To satisfy the observational constraint that the burst spectra are nonthermal, we require $\tau < 1$. By combining equations (8) and (9) with $\alpha = 2$, and imposing the inequality $\tau < 1$, we may obtain a lower bound $R_0 > R_{\rm min}(\gamma, \epsilon_{\rm t})$:

$$R_{\rm min} = \frac{3}{16}\frac{\sigma_{\rm T}}{c}\frac{k\epsilon_{\rm t}}{m_e^2 c^4}\, D^2 f(\gamma), \qquad (10)$$

where $f(\gamma)$ is a numerically-calculated function of $\gamma$. In the case where the spectrum extends to infinite energies ($\epsilon_2 \to \infty$), this function is very well fit by a power law, $f(\gamma) = 0.033\gamma^{-2}$; in the case where $\epsilon_2 = 100$ MeV, it is fit to $\sim 15\%$ accuracy by $f(\gamma) = 0.032\gamma^{-1.79}[1-(\gamma/97)]^{3.07}$. Figure 2 shows the numerically-calculated values of $R_{\rm min}$ for a burst at a distance $D = 1.2$ Gpc, for the two cases. Note that for any value of $\gamma$, the $\gamma$-ray source must be in thermal equilibrium at radii smaller than $R_{\rm min}(\gamma)$.

One needs to relate $R_0$ to observable quantities in order to obtain a useful constraint on $\gamma$ from equation (10). Such a relation is obtained from the comment made above that any variation in the observed burst flux must be longer than the geometric time delay between photon paths (cf. §2.2). Taking into account the cosmological time dilation, $T \gtrsim R_0(1+z)/(2c\gamma^2)$. This in turn gives an upper bound on $R_0$,

$$R_{\max} = \frac{2c\gamma^2 T}{1+z}, \tag{11}$$

where $T$ will be taken to be the observed duration of peak flux emission for the burst. Equations (10) and (11) lead to a lower bound on the Lorentz factor of the expanding material, $\gamma > \gamma_{\min}(\epsilon_t) \equiv [(1+z)R_{\min}/(2cT)]^{1/2}$ (see Fig. 2). For the case where $\epsilon_2 \to \infty$, one can solve analytically for $\gamma_{\min}$ to obtain

$$\gamma_{\min} = \left[ \frac{3}{32} \frac{\sigma_{\mathrm{T}}}{c^2} \frac{k\epsilon_t}{m_e^2 c^4} \frac{0.033(1+z)D^2}{T} \right]^{1/4}. \tag{12}$$

In the case where $\epsilon_2 = 100$ MeV, equations (10) and (11) must be solved numerically for $\gamma_{\min}$ by Newton-Raphson iteration. As we will see in §3.1, the quantity $kD^2$ is independent of the observed properties of a given burst.

Some GRB models have an unavoidable amount of baryonic matter present, such as those involving neutron stars (Woosley 1993; Eichler et al. 1989; Mészáros & Rees 1992; Narayan, Paczyński, & Piran 1992; Usov 1992; Loeb 1993). Thus, an interesting empirical constraint would be an upper bound on the baryonic contamination allowed for a burst source. The total bulk kinetic energy of the matter is $\gamma M c^2$, where $M$ is the total rest mass. It may be that not all of this energy is converted to $\gamma$-rays, and only a fraction $\xi \equiv E_\gamma/(\gamma M c^2)$ of the total kinetic energy is radiated as $\gamma$-ray energy $E_\gamma$. The maximum baryonic mass allowed is then $M_{\max} = E_\gamma/(\xi \gamma_{\min} c^2)$.

Also of interest is the type of environment the GRB sources inhabit. If bursts occur in galaxies, the density of the surrounding medium should be typical of the interstellar medium ($\sim 1$ cm$^{-3}$). The above bounds on $\gamma$ and $M$ can be used to set an upper limit on the ambient gas density. For a spherical expansion, the minimum amount of matter present is the amount swept up by the expanding fireball, $M_{\mathrm{ISM}} = (4\pi/3)\overline{m} n_{\mathrm{ISM}} R_0^3$, where $\overline{m}$ is the mean atomic mass and $n_{\mathrm{ISM}}$ is the interstellar particle number density. The requirement $M_{\mathrm{ISM}} < M_{\max}$ yields an upper bound on the gas density surrounding the source: $n_{\mathrm{ISM}} < n_{\max} \equiv M/(4\pi R_{\min}^3/3)$, out to $\langle R_0 \rangle \lesssim 10^{16}$ cm. Note that $n_{\max}$ depends on the Lorentz factor $\gamma$ of the source; if GRB spectra extend to infinite energies, then to leading order $n_{\max} \propto \gamma^5$. In §3.1, $n_{\max}$ is calculated for $\gamma = \gamma_{\min}$; however, the strong $\gamma$-dependence considerably weakens the constraint.

### 2.2 Derivation of the Maximum Curvature Radius of the Emitting Surface

Consider GRB emission from a surface which moves outward with a bulk Lorentz factor $\gamma \gg 1$. The emission is beamed into cones of half-angle $\gamma^{-1} \ll 1$. The surface need not be spherical; the emission could be coming from a collimated jet of opening angle $\gtrsim \gamma^{-1}$ or a part of a more complicated surface. Suppose that a particular burst exhibits significant intensity variations ($\delta C/C \sim 1$) in its count rate $C$ on a timescale $\delta t'$ as measured in the rest (comoving) frame of the expanding fluid. We then postulate that the linear size of the emitting region for this GRB cannot be larger than $c_{\rm sg}\delta t' = \gamma^{-1}c_{\rm sg}\delta t$, where $\delta t = \gamma \delta t'$ is the time interval measured in the lab (nonexpanding) frame, and $c_{\rm sg}$ is the speed at which signals (e.g. sound waves) propagate in the comoving frame. A region of this size is causally disconnected from other such regions on a lab-frame timescale $\delta t$. We assume that there is no modulation of the source which affects all of these regions simultaneously.

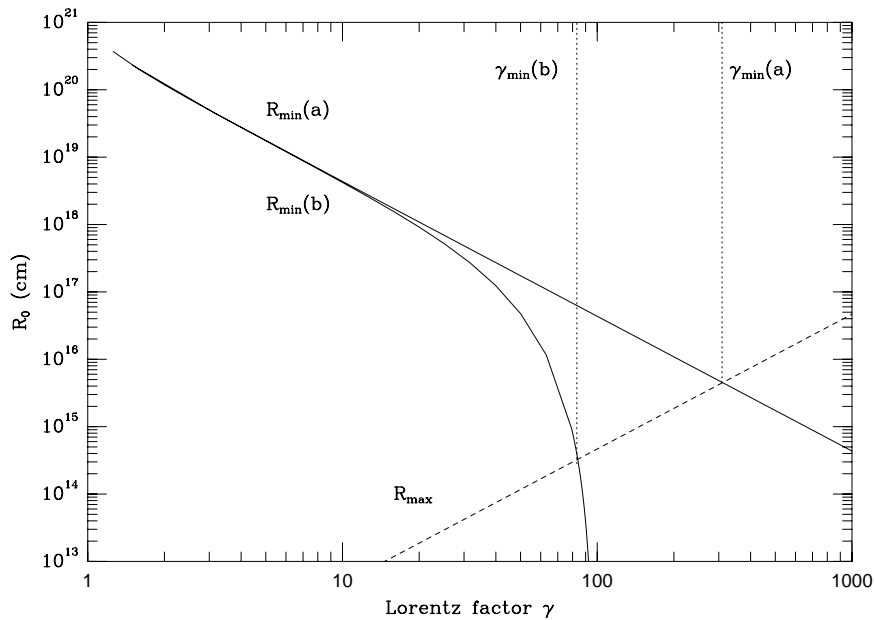

**Fig. 2:** Method of obtaining $\gamma_{\min}$ for a burst. Equation (10) is used to obtain $R_{\min}(\gamma)$ for (a) spectra extending to infinite energies (upper solid curve), and (b) truncated above 100 MeV (lower solid curve). Equation (11) is used to obtain $R_{\max}(\gamma)$ (dashed curve). The intersection of $R_{\min}$ and $R_{\max}$ yields $\gamma_{\min}$ for the two cases. The curves shown are for a burst with an observed duration $T = 1$ sec, for a test photon energy $\epsilon_t = 100$ MeV. The value of $R_{\min}$ is independent of the burst's peak flux (§3.1).

Suppose that the emitting surface has a radius of curvature $R$ with respect to the source (see Fig. 3). A region of solid angle $d\Omega \ll 1$ with respect to the center of curvature then

**Fig. 3:** Geometry for the calculation of the maximum curvature radius $R_{\max}$. After a time $\delta t$, the radiation which was emitted at the beginning of this period has moved ahead of the matter by a distance $(1-\beta)c\delta t$. The causally connected regions are shaded.

contains $dN$ effectively independent emitting sites, where $dN$ is given by the inequality

$$dN > \left(\frac{\gamma R}{c_{\text{sg}}\delta t}\right)^2 \cdot \eta \, d\Omega, \tag{13}$$

and $\eta$ is the fraction of the surface covered by the emitting "cells".

The emitting region seen by a distant observer can be divided into circular rings centered on the line of sight. A ring of angular size $\psi$ (with respect to the center of curvature) and width $d\psi$ contains $N_\psi \equiv (dN/d\Omega) \cdot 2\pi \sin\psi \, d\psi$ emitting cells, where $dN/d\Omega$ is given by equation (13). Due to relativistic beaming, the contribution to the total mean count rate $C$ from this ring is $C_\psi \equiv (dC/d\Omega) \cdot 2\pi \sin\psi \, d\psi$, where

$$\frac{dC}{d\Omega} = \frac{1}{4\pi\gamma^2}\frac{C}{(1-\beta\cos\psi)^2}. \tag{14}$$

(cf. Rybicki & Lightman 1979); $\gamma \gg 1$ is the bulk Lorentz factor, and $\beta \equiv (1-\gamma^{-2})^{1/2}$. For simplicity, the cells may simply be treated as being in either an "on" state or an "off" state during this time. The emission from the ring is then governed by the Poisson statistics of the independent causally disconnected cells, so that the variance of the intensity is $(\delta C)^2_\psi = C^2_\psi/N_\psi$. The variances of all the rings add in quadrature, so the total variance is

$$(\delta C)^2 = \int_0^{\pi/2} \left(\frac{dC}{d\Omega}\right)^2 \left(\frac{dN}{d\Omega}\right)^{-1} \cdot 2\pi \sin\psi \, d\psi. \tag{15}$$

The quantity $dN/d\Omega$ can be treated as a constant (cf. Eq. (13)), and may be pulled outside of the integral. Combining equations (13), (14), and (15) and performing the integration yields an upper bound for the curvature radius $R < R_{\max}$, where

$$R_{\max} = \frac{c_{\rm sg}\delta t}{\sqrt{3\pi\eta}} \frac{C}{\delta C}. \tag{16}$$

The lab-frame emission timescale $\delta t$ is related to the observed variability timescale $\delta t_o$ by

$$\delta t_o = (1+z)\left(\frac{\delta t}{2\gamma^2} + \frac{R}{2c\gamma^2}\frac{\delta C}{C}\right). \tag{17}$$

The first term is due to the velocity difference $c(1-\beta) \approx c/(2\gamma^2)$ between the expanding material and the photons (cf. Fig. 3), and the second term is the geometric time delay given by equation (11). The factor of $\delta C/C$ is necessary because only the fraction $\delta C/C$ of the photons which contribute to the observed fluctuation suffer this geometric delay. The factor $(1+z)$ accounts for time dilation in the case of cosmological sources. In terms of observed quantities, equation (16) becomes

$$R_{\max} = \frac{2\gamma^2 c_{\rm sg}}{\sqrt{3\pi\eta} + (c_{\rm sg}/c)}\frac{\delta t_o}{1+z}\frac{C}{\delta C}. \tag{18}$$

Notice that, to within a factor of order unity, equation (18) resembles equation (11) under the substitutions $T \sim \delta t_o/(\delta C/C)$ and $c_{\rm sg} \sim c$. However, while equation (11) was based on a geometric time delay, equation (18) is based on causality and could in principle yield stricter bounds on $R_{\max}$ if $c_{\rm sg} \ll c$. The method for obtaining $\delta t_o/(\delta C/C)$ from the BATSE light curve data, as well as the problems involved in constraining $\gamma$, are discussed in the next section.

### 3. CONSTRAINTS ON BURST PROPERTIES: RESULTS

In this section we apply the constraints derived in §2 to the BATSE burst population. The methods of extracting the relevant information from the data are discussed, as well as the assumptions made about the population of burst sources.

#### 3.1 Distributions in $E_\gamma$, $\gamma_{\min}$, $M_{\max}$, and $n_{\max}$

The goal of §2.1 was to derive the values of $\gamma_{\min}$, $M_{\max}$, and $n_{\max}$, and to express them in terms of observable quantities. We have already assumed that the burst spectra can be approximated by a power law with spectral index $\alpha = 2$ (Eq. (9)). With this assumption in mind, the three quantities to be obtained from the BATSE data are the distance to the source $D$, the duration of peak emission $T$, and the spectral constant $k$.

To obtain the distance to a burst, one must make assumptions about both the luminosity function of the bursts, and the cosmological model which describes the geometry of the universe. The GRB number count data have been shown to be consistent with a nonevolving population of standard candles in peak luminosity, all with identical power-law spectra $\phi_o \propto \epsilon^{-2}$, in an $\Omega = 1$, $\Lambda = 0$ universe (Mao & Paczyński 1992; Piran 1992). In principle, one might also consider bursts as standard candles in total *energy* and use fluence as a distance indicator; however, because BATSE is sensitive to peak flux, this introduces a significant selection bias against faint, long bursts. For $\Lambda = 0$, the statistical significance of the number-versus-peak-flux fit depends only very weakly on the value of $\Omega$ (Wickramasinghe et al. 1993). We therefore adopt the simplest model and assume a nonevolving population of standard candles with a spectral index $\alpha = 2$ in an $\Omega = 1$ universe. For a given value of $\alpha$, the observed peak number-flux of photons $C_{\text{peak}}(z)$ from a source at a redshift $z$ is given by

$$C_{\text{peak}} = \int_{\epsilon_L}^{\epsilon_U} \phi(\epsilon_o) d\epsilon_o = \frac{\Gamma \cdot (1+z)^{2-\alpha}}{4\pi D^2}, \tag{19}$$

where $\epsilon_L < \epsilon_o < \epsilon_U$ is the energy range picked up by the detector, $\Gamma$ is the intrinsic peak emission rate of photons per unit time of the source in this energy range, and $D$ is its luminosity distance (Weinberg 1972),

$$D = \frac{2c}{H_0}[(1+z) - \sqrt{1+z}]. \tag{20}$$

There are two kinds of peak flux data listed in the BATSE catalog: the peak flux $C_{\text{peak}}$ in photons cm$^{-2}$ sec$^{-1}$, and the ratio of maximum to threshold count rates $C_{\max}/C_{\min}$. We used $C_{\text{peak}}$ as our distance indicator, since this quantity was corrected for detector orientation and atmospheric scattering. Values of $C_{\text{peak}}$ are given for three integration times: 64, 256, and 1024 msec. The value $C_{\text{peak}}^{64}$ in the 64-msec channel is likely to be the best measure of the true peak flux. Due to detector inefficiency at low fluxes ($\lesssim 1$ cm$^{-2}$ sec$^{-1}$; Fishman et al. 1994), only bursts with $C_{\text{peak}} \equiv C_{\text{peak}}^{64} > 1$ cm$^{-2}$ sec$^{-1}$ were included in the sample. To determine the value of $\Gamma$ in equation (19), a one-distribution Kolmogorov-Smirnov (K-S) test was performed on the number-versus-peak-flux distribution for the 254 bursts with $C_{\text{peak}} > 1$ cm$^{-2}$ sec$^{-1}$ and no data gaps (see below). The best-fit value was $\Gamma = 1.0^{+1.7}_{-0.6} \times 10^{57} h^{-2}$ sec$^{-1}$, where the upper and lower bounds denote the 95% confidence region, and $h \equiv H_0/(100$ km sec$^{-1}$ Mpc$^{-1}$) is the dimensionless Hubble constant. Figure 4a shows the fit obtained for the number counts vs. peak flux with this best-fit value of $\Gamma$. Taking this value, the luminosity distance to a standard-candle source with peak flux $C_{\text{peak}}$ is given by equation (19). Figure 4b shows the K-S significance for standard candles as a function of $\Gamma$, indicating that there is actually a range of values which give a good fit. It should also be noted that the significance of the K-S fit was insensitive to the width of the burst luminosity function, as already pointed out by Lubin & Wijers (1993), Ulmer & Wijers (1994), and Cohen & Piran (1995). A luminosity function with a width of the order of its mean is not ruled out, and in this case a value of $C_{\text{peak}}$ would correspond not to a single value of $D$ but rather to a probability distribution in values of $D$. Figure 4c shows the K-S significance as a function of "width" $\sigma$ for a log-normal luminosity function peaked at $\Gamma = 1.0 \times 10^{57} h^{-2}$ sec$^{-1}$ (see Eq. (25)).

The spectral constant $k$ is obtained by assuming that the spectral shape (9), with $\alpha = 2$, applies during the period of peak emission, and using the first equality in equation (19). The BATSE peak flux data are for the channels 50-300 keV, so setting $\epsilon_L = 50$ keV and $\epsilon_U = 300$ keV, one has

$$k = (0.06 \ C_{\text{peak}}) \text{ MeV cm}^{-2} \text{ sec}^{-1}. \tag{21}$$

Note that since $k \propto C_{\text{peak}}$ and $D^2 \propto (C_{\text{peak}})^{-1}$, the optical depth, and hence $\gamma_{\min}$, is independent of $C_{\text{peak}}$ to within a factor of order unity (cf. Eqs. (10)-(12)). This reflects the assumption that GRB are standard candles in peak flux, and hence all have the same photon density at the source.

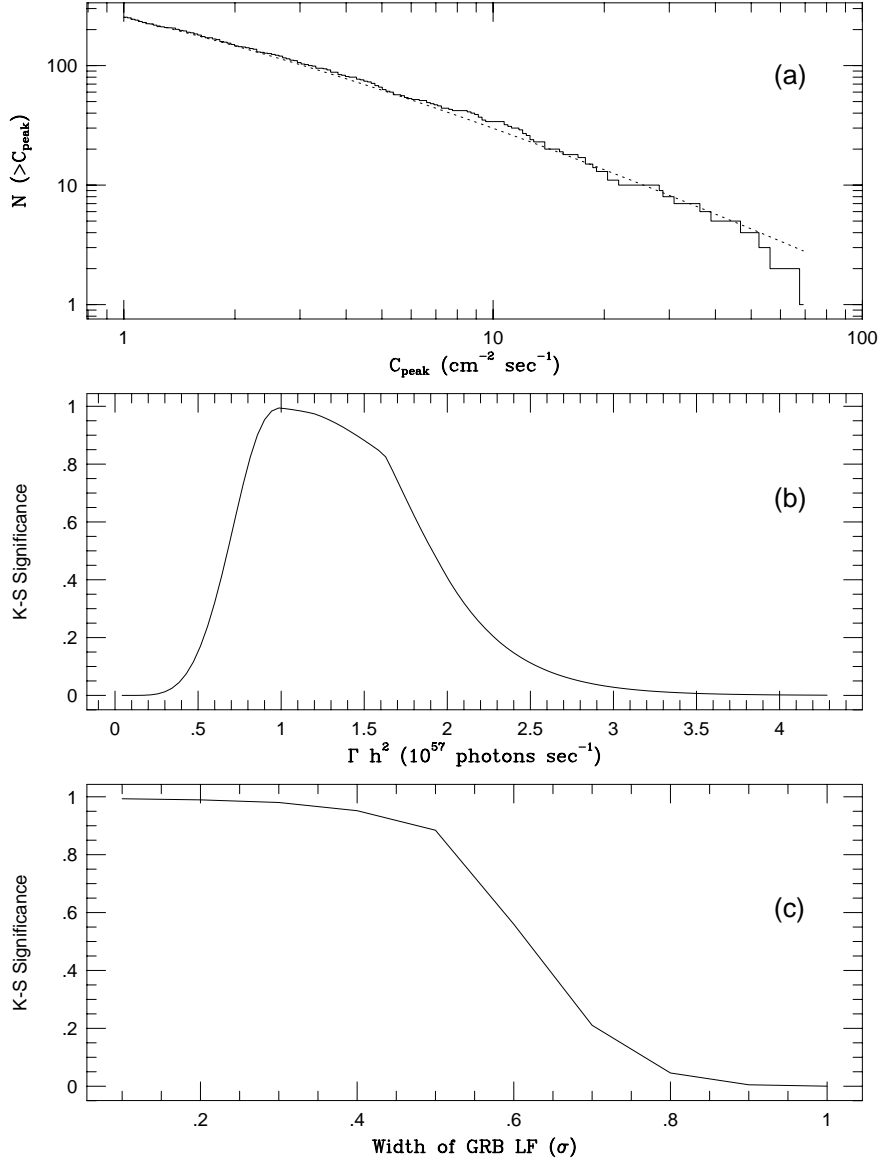

Fig. 4: (a) Number count $N$ vs. peak flux $C_{\rm peak}$ for the 254 bursts in the second BATSE catalog with $C_{\rm peak} > 1$ cm$^{-2}$s$^{-1}$ and no data gaps (solid line), and the best-fit curve for standard candles in an $\Omega = 1$, $h = 0.75$ universe (dotted). (b) Significance of the K-S fit as a function of peak emission rate $\Gamma$ near the best-fit value of $\Gamma = 1.0 \times 10^{57} h^{-2}$ sec$^{-1}$, with $\Omega = 1$ and $h = 0.75$. (c) Significance of the K-S fit as a function of width $\sigma$ for a log-normal luminosity function peaked at $\Gamma = 1.0 \times 10^{57} h^{-2}$ sec$^{-1}$ (Eq. (25)), with $\Omega = 1$ and $h = 0.75$.

The peak emission duration $T$ was formerly expressed only in terms of an upper bound, given by the total burst fluence (time-integrated flux) divided by the peak flux (Woods & Loeb 1994). However, recently the burst lightcurves have become available (Fishman et al. 1994), and so it has been possible to extract $T$ more directly. Here we define $T$ to be the full width at half maximum of the peak, i.e., the time interval from the point when the flux rises above half of its maximum value to the point when it falls below half of the maximum value. In the case of a burst with many peaks, the value of $T$ was computed only for the highest peak. Because $\gamma_{\min}$ depends only weakly on $T$ ($\gamma_{\min} \propto T^{-1/4}$ for spectra extending to infinite energies and even weaker for a finite energy cutoff), the uncertainty introduced in multiply-peaked bursts is small. The method used for determining and subtracting the background is discussed in §3.2. Bursts with data gaps (i.e., where the count rate drops suddenly to zero for some length of time due to instrumental error) were excluded from the sample.

Figure 5 shows the results of applying the constraints derived in §2.1, along with equations (19)-(21), to the 254 bursts in the second BATSE catalog with $C_{\text{peak}} > 1$ cm$^{-2}$ sec$^{-1}$ and no data gaps. The distributions in the minimum Lorentz factor $\gamma_{\min}$, the maximum baryonic mass $M_{\max}$, and maximum ambient density $n_{\max}$ are shown, in addition to the distribution in total $\gamma$-ray energy for a spherical source ($E_\gamma = 4\pi D^2 S$, where $S$ is the burst fluence). The values of $E_\gamma$, $M_{\max}$, and $n_{\max}$ should be multiplied by $\Delta\Omega/4\pi$ for jets that cover a solid angle $\Delta\Omega$. We have taken the observed test photon energy to be $\epsilon_t/(1+z) = 100$ MeV, since this is of the order of the highest photon energy to which BATSE is sensitive. We show distributions for the case where the burst spectra extend to infinite energies (Fig. $5a - d$), and for the case where the spectrum cuts off at 100 MeV (Fig. $6a - d$). Each histogram is normalized to unit area, and shows the fractional number of bursts per logarithmic interval in a given quantity $x$. Thus, each histogram is a discrete plot of $xP(x)$, where $P(x)$ is the probability distribution in $x$ for the burst sample. For spectra extending to infinite energies (curve ($a$) in Fig. 2), one has the leading-order proportionalities $E_\gamma \propto h^{-2}$, $\gamma_{\min} \propto h^{-1/2}\epsilon_t^{1/4}$, $M_{\max} \propto h^{-3/2}\epsilon_t^{-1/4}$, and $n_{\max} \propto h^{3/2}\epsilon_t^{-7/4}$. Note that since $\gamma_{\min}$ is independent of $C_{\text{peak}}$ to within a factor of order unity, the distribution in $\gamma_{\min}$ simply reflects the distribution in peak emission duration $T$. Also note that only $n_{\max}$ is strongly sensitive to the test photon energy $\epsilon_t$; the values of $n_{\max}$ would be lowered appreciably for test photon energies much higher than 100 MeV. However, a modest increase in $\gamma/\gamma_{\min}$ can compensate for this reduction since, to leading order, $n_{\max} \propto \gamma^5$. For the case of an energy cutoff at 100 MeV (curve ($b$) in Fig. 2), the dependences on $\epsilon_t$ and $h$ are weaker. The narrow distribution for $\gamma_{\min}$ in figure 6$b$ can be understood by inspecting figure 2. One obtains only a small range of $\gamma_{\min}$ values by sliding the dashed curve for $R_{\max}$ up and down and changing its point of intersection with the lower solid curve for $R_{\min}(b)$.

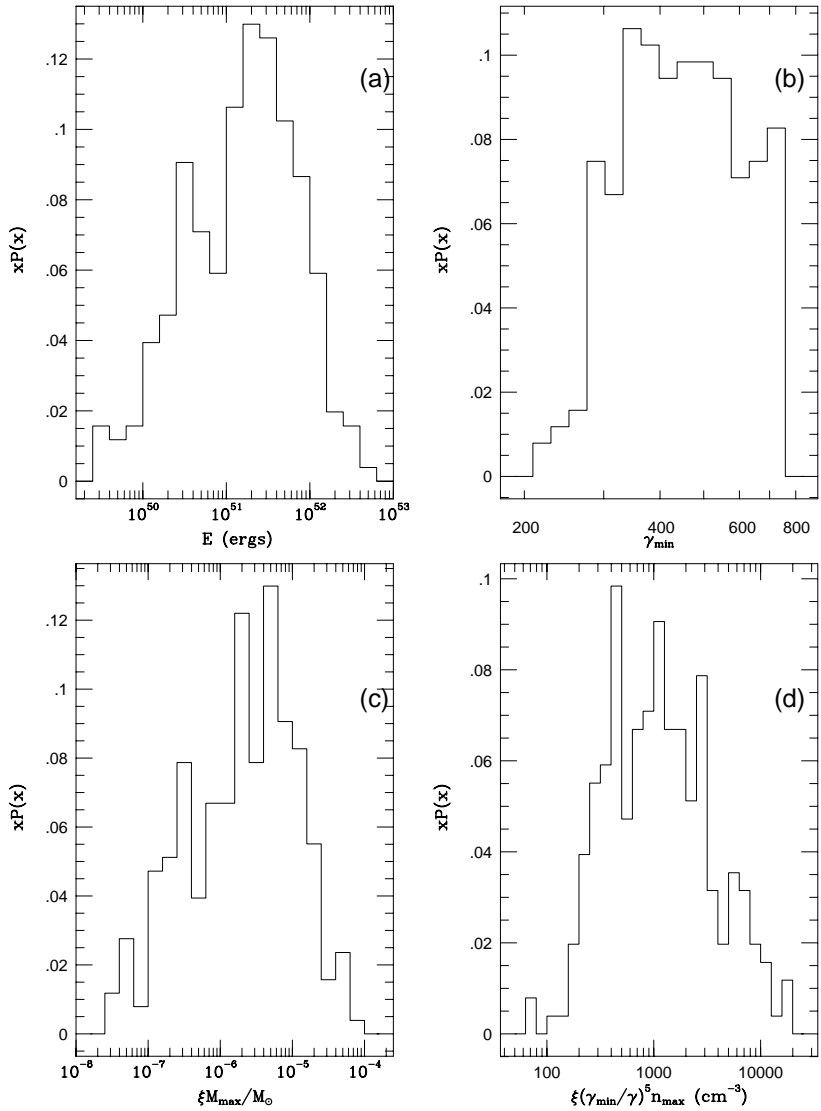

**Fig. 5:** Probability distributions $P(x)$ for ($a$) the radiation energy $E_\gamma$, ($b$) the minimum Lorentz factor $\gamma_{\min}$, ($c$) the maximum baryonic mass $M_{\max}$, and ($d$) the upper bound on the ambient gas density $n_{\max}$, for the case where GRB spectra extend to infinite energies. The histograms show $xP(x)$ for the variable $x$ whose logarithm appears on the horizontal axis. We use a test photon energy $\epsilon_{\rm t} = 100(1+z)$ MeV. Note that only $n_{\max} \propto \epsilon_{\rm t}^{-7/4}$ is significantly sensitive to $\epsilon_{\rm t}$.

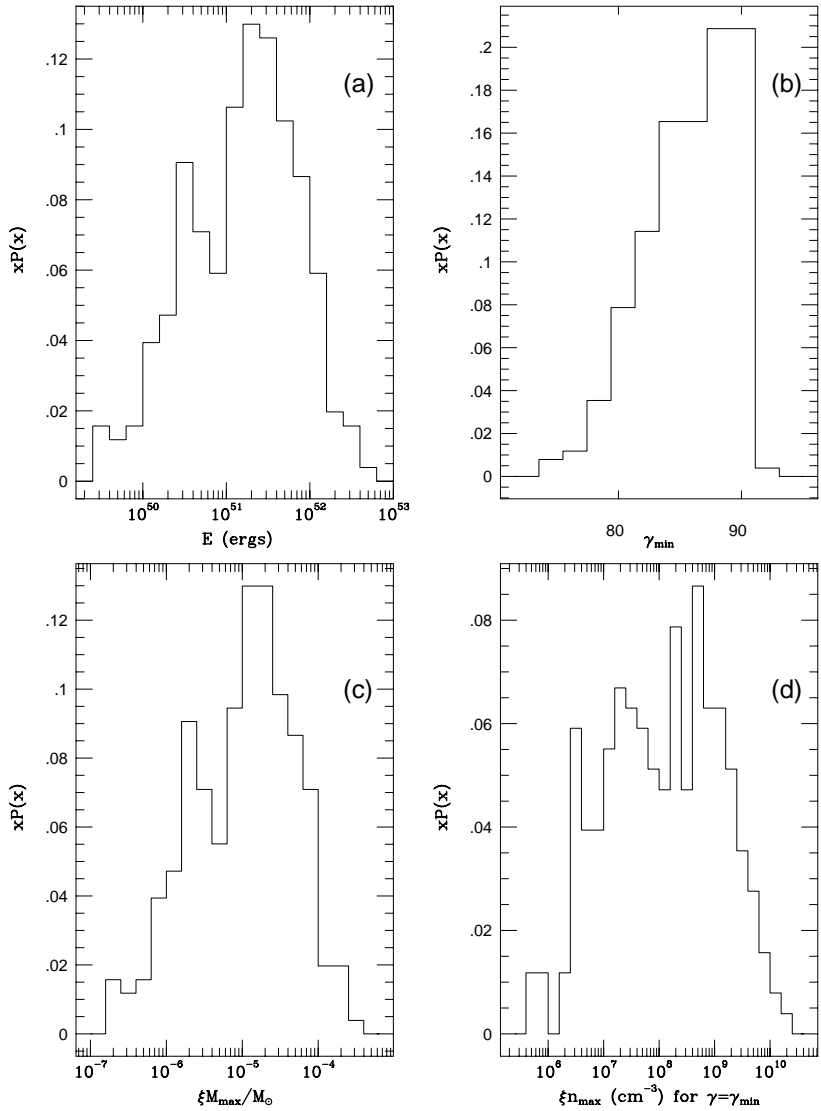

**Fig. 6:** The same as in figure 5, for the case where GRB spectra truncate above 100 MeV. The dependence of $n_{\max}$ on $\gamma$ in this case is more complicated than in figure 5.

Table 1 gives the mean $\langle x \rangle$, the standard deviation $\Delta x = \langle (x - \langle x \rangle)^2 \rangle^{1/2}$, and the log-skewness $s_{\log} = \langle (\log x - \langle \log x \rangle)^3 \rangle / \langle (\log x - \langle \log x \rangle)^2 \rangle^{3/2}$ for the distributions shown in figures 5 and 6. A rough estimate of the significance of the skewness is obtained by comparing it to $(15/N_{\rm tot})^{1/2} \approx 0.24$, where $N_{\rm tot} = 254$ is the total number of data points (Press et al. 1992). All of the distributions are negatively skewed in log space, but not very significantly; they are well approximated by a log-normal profile.

| Moment | $E_\gamma/{\rm erg}$ | $\gamma_{\rm min}$ | $\xi M_{\rm max}/M_\odot$ | $\xi n_{\rm max}/{\rm cm}^{-3}$ |
|---|---|---|---|---|
| Mean | $3.9 \times 10^{51}$ | $4.6 \times 10^2$ | $0.61 \times 10^{-6}$ | $2.1 \times 10^3$ |
| Std. Dev. | $5.9 \times 10^{51}$ | $1.4 \times 10^2$ | $1.0 \times 10^{-6}$ | $2.9 \times 10^3$ |
| Log-Skew | -0.33 | -0.05 | -0.32 | -0.16 |

**Table 1a:** Mean, standard deviation, and log-skewness of the distributions given in figure 5, for the case where GRB spectra extend to infinite energies. Values of $\xi n_{\rm max}$ are shown for $\gamma = \gamma_{\rm min}$.

| Moment | $E_\gamma/{\rm erg}$ | $\gamma_{\rm min}$ | $\xi M_{\rm max}/M_\odot$ | $\xi n_{\rm max}/{\rm cm}^{-3}$ |
|---|---|---|---|---|
| Mean | $3.9 \times 10^{51}$ | $8.5 \times 10^1$ | $2.6 \times 10^{-5}$ | $8.8 \times 10^8$ |
| Std. Dev. | $5.9 \times 10^{51}$ | $0.37 \times 10^1$ | $4.0 \times 10^{-5}$ | $2.2 \times 10^9$ |
| Log-Skew | -0.33 | -0.05 | -0.33 | -0.20 |

**Table 1b:** Mean, standard deviation, and log-skewness of the distributions given in figure 6, for the case where GRB spectra cut off at 100 MeV. Values of $\xi n_{\rm max}$ are shown for $\gamma = \gamma_{\rm min}$.

### 3.2 The Maximum Curvature Radius $R_{\max}$

There are now over 500 bursts whose lightcurves are available in the second BATSE catalog. The time histories are given in most cases for $\sim$ 240 sec after triggering, although some have more ($\sim$ 560 sec for some of the later bursts) and some have less. The data are summed over the energy channels 50-300 keV, and are also summed over all triggered detectors. The time resolution is 64 msec.

The quantity $\delta t_o/(\delta C/C)$ in equation (18) may be considered an operational definition for the variability timescale of a burst, i.e. the time it takes for the count rate to change by an amount comparable to its mean. It is not a unique definition, since it depends on the choice of $\delta t_o$ and over which part of the burst lightcurve one places the "time window" of width $\delta t_o$. For our purposes it is in our interest to take $\delta t_o$ as small as possible and choose that part of the burst when $\delta C/C$ (taken over a time $\delta t_o$) is the largest. This gives us the strictest possible constraint in equation (18). The smallest value of $\delta t_o$ is limited by the time-resolution of the detector. Thus, the constraint on $R$ could be improved substantially with better instruments in the future.

The absolute normalization of the count rates is irrelevant since it is only the ratio $\delta C/C$ which appears in equation (18). However, to obtain the mean count rate $C$ it is necessary to determine and subtract the background count rate. This is difficult to do rigorously because of Poisson noise and the fact that there is no systematic way to isolate a stretch of flat background for all of the bursts. We therefore estimate the background by binning the data very coarsely every $\sim$ 6.4 sec, finding the minimum count rate over the entire burst lightcurve, and dividing by 100 to get the DC background count rate $C_{\text{back}}$ for 64-msec binning. This procedure yields an underestimate because of the noise (given by $\delta C_{\text{back}} \approx C_{\text{back}}^{1/2}$), but only a very small underestimate since the background is estimated for a very large binning size.

The noise also poses a problem in determining the maximum value of $\delta C/C$ over the burst for a given $\delta t_o$. In the absence of noise, one could simply take the smallest possible value of $\delta t = 64$ msec. However, the signal-to-noise ratio is roughly proportional to $(\delta t_o)^{-1/2}$, so we expect that as we increase $\delta t_o$, we should improve the purity of the burst signal. We nominally require a minimum signal-to-noise ratio of 5. The signal-to-noise ratio is defined as $S/N \equiv (\delta C)_{\max}/\delta C_{\text{back}}$, where $(\delta C)_{\max}$ is the maximum value of $\delta C$ for a given $\delta t_o$ and $\delta C_{\text{back}} \approx C_{\text{back}}^{1/2}$ is the noise level, given by the square root of the background count rate in photons sec$^{-1}$. We thus choose the smallest value of $\delta t_o$ such that $S/N > 5$. If a $\delta t_o$ of greater than 0.64 sec is needed, the burst is discarded. The effect of varying the minimum $S/N$ is discussed below.

Figure 7a shows the distribution in values of $R_{\max}$ for the 319 bursts in the second BATSE catalog for which $S/N > 5$ with $\delta t_o < 0.64$ sec. The value of $R_{\max}$ for each burst is calculated with the smallest value of $\delta t_o$ which gives $S/N > 5$. More than half of the bursts have $S/N > 5$ for $\delta t_o = 0.064$, at the limit of BATSE's resolution. This, along with the fact that $\delta C/C < 1$ by definition, accounts for the "pile-up" of events at low values of $R_{\max}$ in figure 7. Better time resolution will lower the values of $R_{\max}$, thus strengthening our constraints.

Changing the signal-to-noise cutoff can in principle change the shape of the $R_{\max}$ distribution. If the cutoff is increased, the sample size will decrease and bursts with larger $\delta t$ will constitute a larger part of the sample. This will skew the $R_{max}$ distribution toward larger values. However, the shape of the histogram does not change qualitatively (see Fig. 7b), and the range of $R_{max}$ values spanned by the sample does not change.

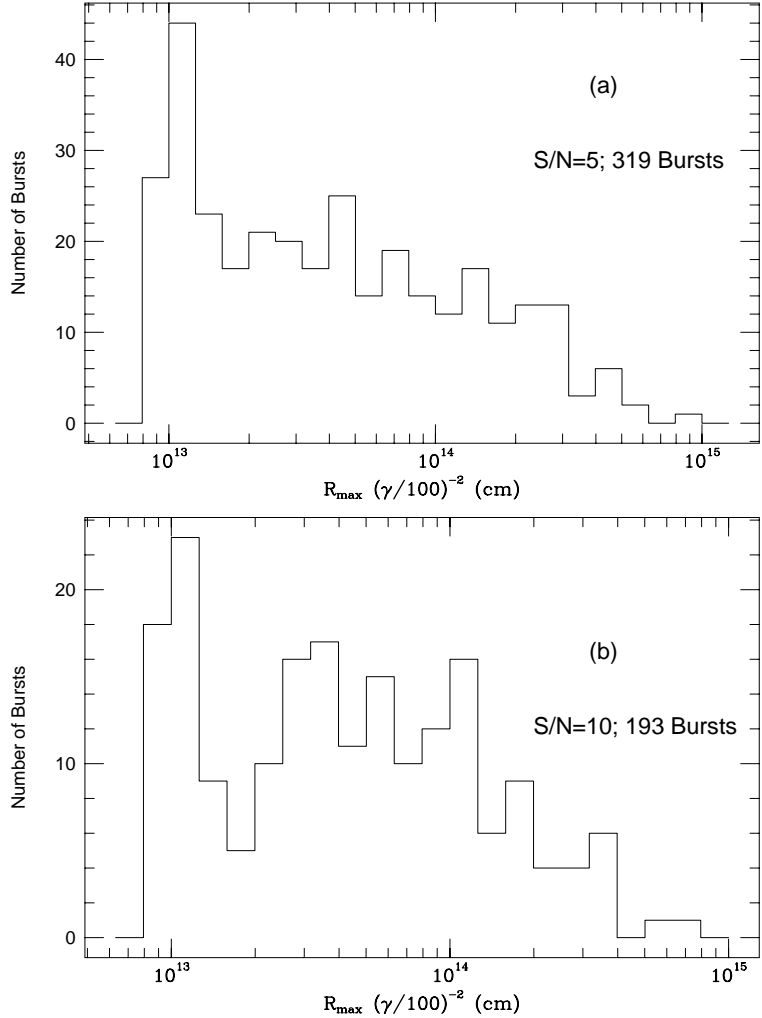

**Fig. 7:** Distribution in values of $R_{max}$, the maximum curvature radius derived in §4, with a minimum signal-to-noise of (a) $S/N = 5$ and (b) $S/N = 10$. Values were calculated for $\gamma = 10^2$, $c_{sg} = c$, and $\eta = 1$ (cf. Eq. (18)).

The biggest uncertainty in figuring the value of $R_{max}$ originates from the unknown value of the Lorentz factor $\gamma$. In principle, one can hope to place an upper bound on $\gamma$. Unfortu-nately, no such bound is strict enough to place a useful limit on $R$. An upper bound on the value of $\gamma$ achievable in a fireball model is the value reached when the Compton drag time exceeds the comoving expansion time in the fireball (Mészáros, Laguna, & Rees 1993). However, this constraint depends on the initial radius of the fireball, which is a model-dependent quantity.

In principle, the signal propagation speed $c_{\rm sg}$ can be nonrelativistic. If the average GRB photon is observed to have an energy $\sim$ MeV, its energy in the rest frame of the expanding material is $\sim$ MeV/$\gamma$. Since $\gamma \gtrsim 10^2$ for most bursts, this would be consistent with a nonrelativistic temperature, and hence a nonrelativistic sound speed, in the material rest frame. If $c_{\rm sg} \ll c$, then the constraint in equation (18) is strengthened considerably relative to equation (11).

## 4. LIMITS ON THE ASSOCIATION OF GRB WITH GALAXIES

The empirical evidence for a cosmological origin of $\gamma$-ray bursts is compelling (Mao & Paczyński 1992; Piran 1992; Meegan et al. 1993; Norris et al. 1993). However, one important ingredient is missing: no host galaxies have been found near any well-localized bursts (Schaefer 1992; Fenimore et al. 1993). In this section we make the assumption that bursts occur in galaxies, and place constraints on the luminosities of the host galaxies and on the width of the GRB luminosity function.

For simplicity, we assume non-evolving populations of both galaxies and $\gamma$-ray burst sources and a flat ($\Omega = 1$) universe. The consequences of introducing galaxy evolution are discussed in §5. If $\gamma$-ray bursts are associated with neutron stars, there is no reason to expect the shape of the GRB luminosity function to evolve with cosmological time, although the comoving density of GRB sources may evolve.

Suppose that the positional error box for a given burst is searched for galaxies in the wavelength band $\lambda$, and that no galaxies are found with an apparent magnitude brighter than $m$. Then, the host galaxy for this burst cannot be brighter than $L_{\rm max}$, where

$$L_{\rm max} = 10^{10} L_\odot \left[\frac{D(z_b)}{10 \text{ pc}}\right]^2 10^{0.4[M_{10} - m + K_\lambda(z_b)]}. \tag{22}$$

Here, $M_{10}$ is the absolute magnitude of a $10^{10} L_\odot$ galaxy ($M_{10} = -19.52$ and $-20.69$ for the $B$ and $R$ bands, respectively), $D(z_b)$ is the luminosity distance to the burst source at redshift $z_b$, and $K_\lambda(z)$ is the $K$-correction for a galaxy at redshift $z$ in the wavelength band $\lambda$. The $K$-correction simply accounts for the redshifting of photons from shorter wavelengths in the galaxy spectrum into the observed band. The correction due to extinction by dust in the Milky Way is assumed to be small. The values of $z_b$ are calculated by assuming that bursts are standard candles in peak flux and using equations (19) and (20), with the best-fit peak emission rate $\Gamma = 1.0 \times 10^{57}$ sec$^{-1}$. The peak flux values were obtained from Fenimore et al. (1993).

Table 2 lists Schaefer's limiting magnitude and band, our standard-candle burst redshift $z_b$, the values of $L_{\rm max}$, the size $\Delta\Omega$ of the error box in arcmin$^2$ on the sky, and the number of galaxies $N_{\rm gal}$ brighter than the limiting magnitude that one would expect to see in a randomly placed field the size of the error box, for eight of the PVO (Pioneer Venus Orbiter) bursts examined by Schaefer (1992). Values of $L_{\max}$ are given separately for elliptical and spiral galaxies, since the $K$-correction depends on the galaxy type. The $K$-corrections were obtained by interpolating between the redshift data points given by Frei and Gunn (1994). Our values of $z_b$ are systematically higher than those obtained by Fenimore et al. (1993) by $\Delta z_b \sim 0.05$, since our best-fit luminosity and assumed spectral shape for GRB are different. The expected number of galaxies was calculated for the $B$ band from the data of Lilly (1993). The numbers for the $R$ band were calculated using a no-evolution model. The sizes of the positional error boxes were obtained from the catalog of Atteia et al (1987). Note that the nondetection of galaxies near the bursts of 24 Nov 1978 and 6 Apr 1979 is statistically highly unlikely, and so we find it surprising that no galaxies were seen.

| Burst | $m$ (band) | $z_b$ | $L_{\max}$(Ell.) ($10^{10} L_\odot$) | $L_{\max}$(Sp.) ($10^{10} L_\odot$) | $\Delta\Omega/(1')^2$ | $N_{\text{gal}}$ |
|---|---|---|---|---|---|---|
| 19 Nov 78 | 16.3 ($B$) | 0.11 | 17.2 | 13.3 | 8. | 0.01 |
| 24 Nov 78 | 20.0 ($R$) | 0.13 | 0.18 | 0.17 | 48. | 6. |
| 25 Mar 79 | 20.0 ($R$) | 0.19 | 0.42 | 0.37 | 2 | 0.2 |
| 29 Mar 79 | 18.0 ($B$) | 0.25 | 31.0 | 20.3 | 41. | 0.2 |
| 06 Apr 79 | 24.7 ($B$) | 0.17 | 0.02 | 0.02 | 0.26 | 3. |
| 18 Apr 79 | 20.5 ($B$) | 0.14 | 0.57 | 0.43 | 2.9 | 0.3 |
| 13 Jun 79 | 19.5 ($R$) | 0.27 | 1.51 | 1.25 | 0.76 | 0.05 |
| 16 Nov 79 | 21.0 ($B$) | 0.11 | 0.21 | 0.17 | 3.7 | 0.7 |

**Table 2:** Limiting magnitude $m$ (Schaefer 1992), source redshift $z_b$ (calculated using Eq. (19)-(20)), limiting luminosity $L_{\max}$ in units of $10^{10} L_\odot$ (cf. Eq. (22)), size $\Delta\Omega$ of the positional error box in arcmin$^2$, and expected number of galaxies $N_{\text{gal}}$ in the burst error box, for eight well-localized PVO bursts. The values of $z_b$ and $L_{\max}$ are based on the assumption that GRB are standard candles in peak flux. Values are shown separately for elliptical (Ell.) and spiral (Sp.) galaxies because of the different $K$-corrections for each type.

If GRB do occur in galaxies, one may calculate the probability $P$ of *not* seeing the host galaxy for a given burst in a wavelength band $\lambda$ down to a limiting magnitude $m$, or equivalently, the fraction of the total number of potential host galaxies which are intrinsically fainter than the limiting luminosity $L_{\max}$ given by equation (22). Again we assume no evolution and take the luminosity functions for each of the galaxy types to be Schechter functions, so that the number galaxies per Mpc$^3$ in the luminosity interval $(L, L + dL)$ at a redshift $z$ is given by

$$\phi_i(z,L)\, dL = (1+z)^3 \phi_{*,i} \left(\frac{L}{L_{*,i}}\right)^{\alpha_i} \exp\left(-\frac{L}{L_{*,i}}\right) \frac{dL}{L_{*,i}}, \tag{23}$$

where $i$ refers to the morphological type and $\phi_{*,i}$, $L_{*,i}$, and $\alpha_i$ for each type taken from the fits to the CfA redshift survey data (Marzke, Geller, & Huchra 1994). These fits were obtained using Zwicky magnitudes; we use the same fitted values of $\phi_{*,i}$, etc. for the $B$, $V$, and $R$ bands probed by Schaefer (1992). The factor of $(1+z)^3$ in equation (23) accounts for the scaling of the proper density of galaxies with cosmological redshift. Suppose that GRB only occur in galaxies brighter than $L_{\min}$. If bursts are standard candles in peak flux, then the desired probability is

$$P = \frac{\sum_i \int_{L_{\min}}^{L_{\max,i}(z_b,m)} dL \; \phi_i(z,L)}{\sum_i \int_{L_{\min}}^{\infty} dL \; \phi_i(z,L)}, \qquad (24)$$

where $L_{\max,i}(z_b)$ is given by equation (22) and $z_b$ is the burst redshift. For standard candles, $z_b$ can be determined from equations (19) and (20). All of the bursts considered here are relatively bright, and have standard-candle redshifts $z_b \lesssim 0.25$. This justifies our neglect of evolution for the galaxy luminosity function. At such low redshifts, one might also expect a low value of $P$, i.e., most galaxies should be visible down to a limiting magnitude as faint as 24. However, bursts are not necessarily standard candles, and the width of the GRB luminosity function has not been well-constrained (Lubin & Wijers 1993; Ulmer & Wijers 1994; Cohen & Piran 1995). In principle, the probability that apparently bright bursts originate at large distances increases as the width of the GRB luminosity function increases. It is therefore conceivable that above a certain width of the GRB luminosity function, the probability for non-detection of the distant host galaxies would be considerable.

To quantify the width of the GRB luminosity function over a large range of values we use a "log-normal" distribution. The probability that a given burst falls in the interval $(\Gamma, \Gamma + d\Gamma)$ is then

$$p_b(\Gamma) \, d\Gamma = \frac{e^{-\sigma^2/2}}{\sqrt{2\pi\sigma^2}} \exp\left\{\frac{-[\ln(\Gamma/\Gamma_0)]^2}{2\sigma^2}\right\} \frac{d\Gamma}{\Gamma_0}, \qquad (25)$$

where $\Gamma$ is the intrinsic peak emission rate in photons sec$^{-1}$ and $\Gamma_0$ is the best-fit value of $\Gamma$ obtained in §3.1 (cf. Eq. (19)). The value of $p_b(\Gamma)$ in equation (25) is normalized to a unit area. A burst with observed peak flux $C_{\text{peak}}$ then has a probability $p_b(z)dz = p_b(\Gamma)(d\Gamma/dz)dz$ of being in the redshift range $(z, z+dz)$. For an $\Omega = 1$ cosmology, the redshift probability distribution is thus

$$p_b(z) = \frac{2C_{\text{peak}}}{C_0} \frac{e^{-\sigma^2/2}}{\sqrt{2\pi\sigma^2}} \left(1 + z - \frac{3}{2}\sqrt{1+z} + \frac{1}{2}\right) \exp\left\{\frac{-[\ln(\Gamma(z)/\Gamma_0)]^2}{2\sigma^2}\right\}, \qquad (26)$$

where $\Gamma(z) = 16\pi(c/H_0)^2 C_{\text{peak}}(1 + z - \sqrt{1+z})^2$ (cf. Eq. (19) and (20)), and $C_0 \equiv \Gamma_0/[16\pi(c/H_0)^2]$. In this case, the probability in equation (24) is modified to an integral over the burst redshift distribution,

$$P = \frac{\sum_i \int_0^\infty dz \; p_b(z) \int_{L_{\min}}^{L_{\max,i}(z,m)} dL \; \phi_i(L,z)}{\sum_i \int_0^\infty dz \; p_b(z) \int_{L_{\min}}^{\infty} dL \; \phi_i(L,z)}. \qquad (27)$$

Figure 8 shows the probability $P$ that the host galaxies were not seen for six of the eight PVO bursts under consideration. The burst luminosity function (25) was chosen to be centered on the best-fit emission rate $\Gamma_0$ from §3, with a width $\sigma$ ranging between 0.1 and 2.0. Note that this range is larger than that given by good fits to the $N$ vs. $C_{\text{peak}}$ curve in Fig. 4c. The morphological classes for which we obtained our $K$-corrections from Frei & Gunn (1994) were assumed to correspond roughly to the classifications used by Marzke et al. (1994) in determining the galaxy luminosity functions. The probabilities are shown as a function of $\sigma$ for each burst, for the three values $L_{\min} = 10^8 L_\odot$, $10^9 L_\odot$, and $10^{10} L_\odot$. The events of 19 Nov 1978 and 29 Mar 1979 both have $P \approx 1$ even for $\sigma = 0$ (i.e., standard candles), and are therefore not shown. For most of the other bursts, the probability becomes large only for values of $\sigma$ which are inconsistent with the fits in figure 4c. We expect that choosing a GRB luminosity function which gives a better $N$ vs. $C_{\text{peak}}$ fit for large widths will not be sufficiently skewed toward high luminosities to produce a stronger dependence of $P$ on the luminosity function width. The fact that no galaxies were seen in the error boxes for the bursts of 24 Nov 1978 and 6 Apr 1979 is surprising even if one does not assume an association between GRB sources and galaxies. This leads us to question the validity of some of the observed limits.

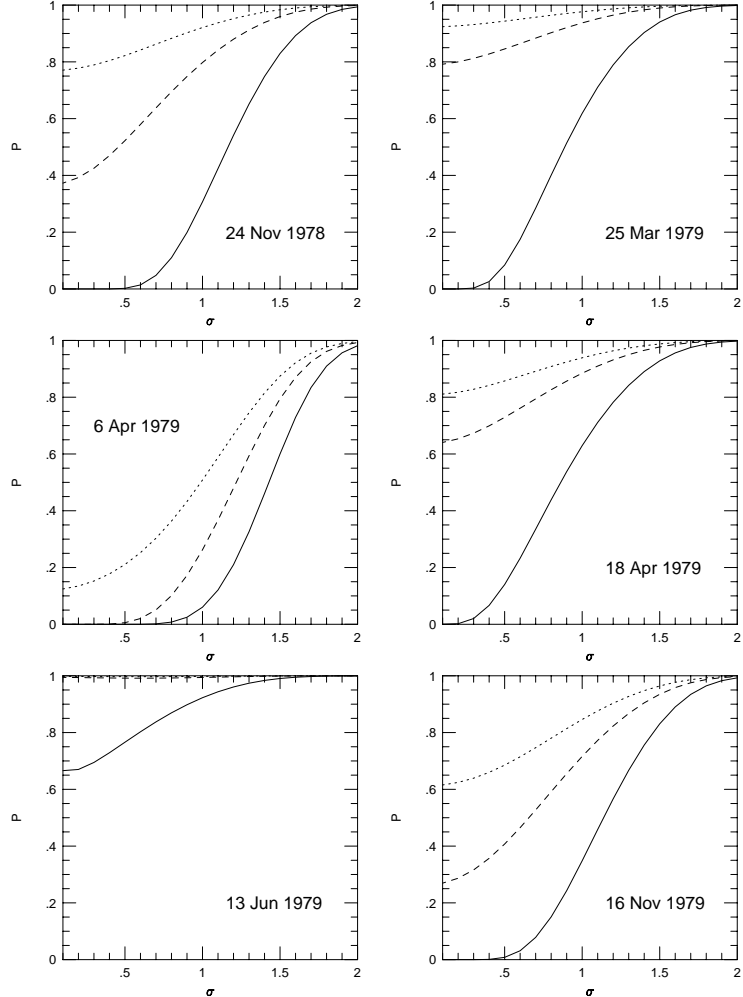

**Fig. 8:** Probability $P$ of not seeing the host galaxy as a function of the width $\sigma$ of the GRB luminosity function (cf. Eq. (25)), for six of the PVO bursts considered by Schaefer (1992). Values are plotted for $L_{\min} = 10^8 L_\odot$ (dotted line), $10^9 L_\odot$ (dashed line), and $10^{10} L_\odot$ (solid line). The bursts of 19 Nov 1978 and 29 Mar 1979 both have $P \approx 1$ for all values of $\sigma$ (see Table 2).

## 5. SUMMARY AND DISCUSSION

In this paper, we have derived several empirical constraints on the physical properties of $\gamma$-ray burst (GRB) sources. In §2.1 and §3.1, the requirement of optical thinness to pair production was used to constrain the bulk Lorentz factors and other physical properties of bursts in the second BATSE catalog. In §2.2 and §3.2, the time variabilities of bursts in the second BATSE catalog were used to place an upper bound on the curvature radius of the emitting surface which produces the burst. In §4, we used the data on several PVO bursts (Schaefer 1992; Fenimore et al. 1993) to examine the likelihood that GRB occur in unseen galaxies as a function of the width of the GRB luminosity distribution.

The results in §2.1 and §3.1 were obtained under four simplifying assumptions. First, GRB were taken to be standard candles in peak luminosity, with $C_{\text{peak}}$ as a distance indicator. In fact, the K-S significance of the $N - C_{\text{peak}}$ fit (Fig. 4c) is not very sensitive to the width of the GRB luminosity function (Lubin & Wijers 1993; Ulmer & Wijers 1994; Cohen & Piran 1995). As discussed in §4 and below, a wide GRB luminosity function may be needed to explain the nondetection of galactic hosts near burst locations on the sky. In this case, a single value of $C_{\text{peak}}$ would correspond to a distribution in redshifts, and each burst would be spread over more than one bin in the histograms of figures 5 and 6. The second assumption was that the peak flux emission of the source is steady over a timescale $\sim R_0/c$. A shorter burst would have a smaller optical depth, because photons emitted at an angle $\sim 1/\gamma$ would be unable to catch up with photons emitted along the line of sight to the center of the source. If this assumption is violated, the constraint on $\gamma$ is weakened. Finally, the geometric time delay expressed in equation (11) was derived assuming that the emission comes from a surface of size $R_0/\gamma$. If the emitting region is a smaller clump of matter, the constraint on $\gamma$ is weakened.

Bearing these assumptions in mind, we arrive at a few physically interesting results from §2.1 and §3.1. The energy distribution in figures 5a and 6a ends at about the binding energy of a neutron star, $\sim 10^{53}$ ergs. This result is consistent with a variety of models that associate cosmological $\gamma$-ray bursts with neutron stars (Eichler et al. 1989; Mészáros & Rees 1992; Narayan, Paczyński, & Piran 1992; Usov 1992; Loeb 1993), although in some models it is difficult to extract all the binding energy in $\gamma$-rays. The distribution of maximum ambient gas densities $n_{\max}$ (Figs. 5d and 6d) allows bursts to reside within the interstellar medium of galaxies, where the density $n_{\rm ISM} \approx 1$ cm$^{-3}$, and is also consistent with a molecular cloud environment ($n \sim 10^{2-4}$ cm$^{-3}$).

The limit $R_{\max}$ derived in §2.2 and §3.2 on the curvature radius of the emitting surface was obtained assuming that the causally disconnected regions on the surface act as truly independent sites. It is possible that many such regions are lit up at the same time due to some overlap in their past light cones, but we ignore this possibility, assuming that the conditions responsible for the emission are local. The fireball model of Mészáros, Laguna, & Rees (1993) places an effective lower bound on the value of $R$ at which the main emission can take place. Since the observed $\gamma$-ray energy must be converted from the bulk kinetic energy of the fireball matter, the main part of the burst should not occur until the expansion of the fireball is decelerated appreciably by the inertia of the surrounding material. This happens when the mass of interstellar matter swept up by the fireball is comparable to the bulk kinetic energy of the baryons in the fireball. For a spherical fireball with a typical energy ($\sim 10^{51}$ ergs) and interstellar ambient density ($\sim 1$ cm$^{-3}$), the radius at which the deceleration becomes significant is $\sim 10^{16}(\gamma/10^3)^{-2/3}$ cm. To reconcile this with the typical value of $R_{\max} = 10^{14}(\gamma/10^2)^2$ found in figure 7, one needs $\gamma \gtrsim 10^3$, a stricter bound than that derived from the optical depth to pair production in §3.1. The alternative is to have emission from collisions between shells in the fireball at smaller radii (Rees & Mészáros 1994). If the signal speed in the local rest frame is much smaller than the speed of light, the constraints on $R_{\max}$ and $\gamma$ become stronger. It should be emphasized that $R_{\max}$ applies to Galactic as well as cosmological sources.

In §4 it was assumed that the luminosity functions of galaxies did not vary with redshift (the "no-evolution" model). The number of galaxies actually observed at blue magnitudes > 20 is larger than the no-evolution prediction. Thus, introducing an evolutionary model which agrees with galaxy counts will only strengthen our result that the probability of not seeing the GRB hosts is low. The corrections due to extinction by dust in the Milky Way were assumed to be small. This is justified for five of the eight bursts considered since these bursts occurred at galactic latitudes $|b| > 55°$, where there is virtually no extinction; for the other three bursts, the extinction correction is not likely to be much more than two-tenths of a magnitude (Burstein & Heiles 1982). The non-detection of galaxies near the bursts of 24 Nov 1978 and 6 Apr 1979 is surprising, given the expected counts based on galaxy surveys. We would expect to find a few galaxies brighter than the limiting magnitude for each of these two bursts in a randomly-placed field the size of the associated error box (Table 2). In the most restrictive case, the event of 6 Apr 1979, the host galaxy cannot be brighter than $\sim 10^8 L_\odot$ if GRBs are standard candles (Table 2). Even if one allows for a GRB luminosity function centered on the best-fit value from §3.1 with a width as large as its mean, the probability of not seeing the host galaxy for this burst is very small unless galaxies with $L \lesssim 10^8 L_\odot$ are more likely to host $\gamma$-ray bursts. Generally, one might expect that the $\gamma$-ray burst rate in a given galaxy would be roughly proportional to the number of stars and hence to the total luminosity. However, if most of the light in the universe originates from faint galaxies, one should see many GRB events occurring in faint host galaxies. In our no-evolution model, the faint-end logarithmic slope of the total galaxy luminosity function is $\alpha \sim -1$ (Marzke et al. 1994), so that the luminosity per logarithmic interval is $L^2\phi(L) \propto L$, and thus faint galaxies do not contribute most of the light. However, an excess of faint galaxies at redshifts $z \gtrsim 0.1$ or a variation in the mass function of stars in faint galaxies may allow for a larger abundance of neutron stars in faint hosts. In addition, there may be a large population of sources which exist outside of galaxies. High-velocity ($\gtrsim 500$ km sec$^{-1}$) neutron star binaries which escape from galaxies could easily wander outside of the 1 arcmin$^2$ box within a Hubble time. There could also be a population of stars formed in intergalactic space. As shown in figure 8, another solution to the problem is to introduce a broad GRB luminosity function. However, figure 4c demonstrates that any such luminosity function is significantly constrained by the number-count statistics of the bursts.

We thank Stirling Colgate, George Field, Josh Grindlay, Shude Mao, Ramesh Narayan, Bohdan Paczyński, and Eli Waxman for insightful comments and suggestions.

## REFERENCES

Atteia, J.-L., et al. 1987, ApJS, 64, 305

Babul, A., Paczyński, B., & Spergel, D. 1987, ApJ, 316, L49

Berestetskii, V. B., Lifshitz, E. M., & Pitaevskii, L. P. 1982, Quantum Electrodynamics (New York: Pergamon), p. 371

Briggs, M. S., et al. 1993, in Proc. of the Huntsville Gamma-Ray Burst Workshop, ed. G. Fishman, J. Brainerd, & K. Hurley (New York: AIP), p. 44

Broadhurst, T. J., Ellis, R. S., & Glazebrook, K. 1992, Nature, 355, 55

Burstein, D., & Heiles, C. 1982, AJ, 87, 1165

Cohen, E., & Piran, T. 1995, ApJL, in press

Dermer, C. D. 1992, Phys. Rev. Lett., 68, 1799

Eichler, D., Livio, M., Piran, T., & Schramm, D. N. 1989, Nature, 340, 126

Fenimore, E. E. et al. 1993, Nature, 366, 40

Fenimore, E. E., Epstein, R. I., & Ho, C. 1993, A&AS, 97, 59

Fishman, G., et al. 1994, ApJS, 92, 229

Frei, Z., & Gunn, J. E., 1994, AJ, 108, 1476

Goodman, J. 1986, ApJ, 308, L47

Hakkila, J., et al. 1994, ApJ, 422, 659

Lilly, S. J. 1993, ApJ, 411, 501

Loeb, A. 1993, Phys. Rev. D, 48, 3419

Lubin, L. M., & Wijers, R. A. M. J. 1993, ApJ, 418, L9

Mao, S., & Paczyński, B. 1992, ApJ, 388, L45

Mao, S., & Paczyński, B. 1992, ApJ, 389, L13

Marzke, R. O., Geller, M. J., & Huchra, J. P., 1994, AJ, in press

Meegan, C. A., et al. 1993, in Proc. of the Huntsville Gamma-Ray Burst Workshop, ed. G. Fishman, J. Brainerd, & K. Hurley (New York: AIP), p. 3

Mészáros, P., Laguna, P., & Rees, M. J. 1993, ApJ, 415, 181

Mészáros, P., & Rees, M. J. 1992, MNRAS, 257, 29P

Narayan, R., Paczyński, B., & Piran, T. 1992, ApJ, 395, L83

Nemiroff, R. J. 1994, Comments on Astrophysics, 17, 4, p. 189

Norris, J. 1993, in Proc. of the Huntsville Gamma-Ray Burst Workshop, ed. G. Fishman, J. Brainerd, & K. Hurley (New York: AIP), p. 177

Paczyński, B. 1986, ApJ, 308, L43

Paczyński, B. 1988, ApJ, 335, 525

Piran, T. 1992, ApJ, 389, L45

Press, W., Teukolsky, S., Vetterling, W., & Flannery, B. 1992, Numerical Recipes (Cambridge: Cambridge Univ. Press), p. 606

Rees, M. J., & Mészáros, P. 1994, preprint astro-ph/9404038

Rybicki, G., & Lightman, A. 1979, Radiative Processes in Astrophysics (New York: Wiley), p. 140-141

Schaefer, B. E. 1992, in Gamma-Ray Bursts: Observations, Analyses and Theories, ed. C. Ho, R. Epstein, & E. Fenimore (Cambridge: Cambridge University Press), p. 107

Schaefer, B. E. 1993, in Proc. of the Huntsville Gamma-Ray Burst Workshop, ed. G. Fishman, J. Brainerd, & K. Hurley (New York: AIP), p. 382

Ulmer, A., & Wijers, R. A. M. J. 1994, ApJ, in press

Usov, V. V. 1992, Nature, 357, 472

Weinberg, S. 1972, Gravitation and Cosmology (New York: Wiley), p. 485

Wickramasinghe, W. A. D. T., et al. 1993, ApJ, 411, L55

Woods, E., & Loeb, A. 1994, ApJ, 425, L63

Woosley, S. E. 1993, ApJ, 405, 273